\documentclass[useAMS,usenatbib]{mn2e}
\usepackage{latexsym,amsmath,amssymb,}
\usepackage{graphicx}

\def\be{\begin{equation}}
\def\ee{\end{equation}}
\def\beq{\begin{eqnarray}}
\def\eeq{\end{eqnarray}}
\def\bes{\begin{eqnarray}}
\def\ees{\end{eqnarray}}

\newlength{\sizeonefig}
\newlength{\sizetwofig}
\title[Confronting models for the HF QPOs with Lense-Thirring precession]{Confronting models for the high-frequency QPOs with Lense-Thirring precession}
\author[I. Zh. Stefanov]{Ivan Zh. Stefanov$^{1}$\thanks{E-mail: izhivkov@tu-sofia.bg}\\
$^{1}$Department of Applied Physics, Technical University of Sofia\\
8, St. Kliment Ohridski Blvd., 1000 Sofia, Bulgaria}
\begin{document}


\pagerange{\pageref{firstpage}--\pageref{lastpage}} \pubyear{2014}

\maketitle

\label{firstpage}

\begin{abstract}
Quasiperiodic oscillations (QPOs) have been observed in the power-density spectra of some low-mass X-ray binaries(LMXB) containing a black hole. The two major groups of QPOs -- low-frequency (LF) and high-frequency (HF)-- have rather different properties. That is why  they are usually studied separately. In the literature one can find a large number of models for the high-frequency QPOs but not so many for the low-frequency ones. HF QPOs have attracted significant research efforts due to their potential to provide indispensable information for the properties of the black hole, for its accretion disc and for strong field gravity in general. However, in order to interpret the data for the HF QPOs of the observed objects we have to fix a model. Here we propose a simple test which could allow us to sift the models. The test is based on five rather general assumptions concerning the nature of the central object in black-hole binaries and the mechanism for the generation of the LF QPOs observed in the PDS of such objects. In other words we combine facts that we know about the LF and the HF QPOs of several objects and search for conflicts. As a result we single out a model for the HF QPOs -- the $3:2$ nolinear resonance model. As a byproduct of this study we propose loose constraints on the mass of the LMXB H 1743-322.
\end{abstract}


\section{Introduction}
Quasiperiodic oscillations (QPOs) have been observed in the power-density spectra of some low-mass X-ray binaries containing a black hole
\citep{Klis_REVIEW,BHBs_McClintock,Review_Zhang}. They have attracted significant research interest since once we find the mechanism responsible for them we will have an indispensable probe of strong field gravity. Two major types of QPOs are identified -- low-frequency (LF) and high-frequency (HF). The former have frequencies ranging (roughly) in the interval $0.1~-–~30$ Hz and have been observed for 14 black-hole binaries. They are strong and persistent, and drift in frequency. 

The HFQPOs, $40 -– 450$ Hz, have been detected for seven sources \citep{BHBs_McClintock,Remillard_Xray_BHBs} 5 black-hole binaries (BHBs) and 2 black-hole candidates (BHCs). Unlike LFQPOs, high-frequency ones are weak and transient. What is probably most important about them is that they do not shift in frequency and that in some sources they occur in pairs with commensurate frequencies -- a lower frequency $\nu_{\rm L}$ and upper frequency $\nu_{\rm U}$, in a $3:2$ ratio.

The difference of the properties of LFQPOs and HFQPOs is an indication that they might be related to different physical processes and
different mechanisms might be responsible for their occurrence.  That is why the two major types of QPOs are usually studied separately. Still, it is a common belief that the reason for both LF and HF QPOs is that the X-ray emission is modulated by the fundamental frequencies of motion of the accreting matter.

The numerical values of the high frequencies are of close magnitude to the fundamental frequencies of motion -- orbital frequency $\nu_{\rm \phi}$, radial $\nu_{\rm r}$ and vertical $\nu_{\rm \theta}$ epicyclic frequencies, of a test particle orbiting along circular geodesics in the inner region of the accretion disc, i.e. close to the innermost stable circular orbit (ISCO). This observation is a motivation for many models to associate the HFQPOs with these frequencies or with simple combinations of them. Each model proposes different mechanism and gives different recipe for the expression of $\nu_{\rm L}$ and $\nu_{\rm U}$ by  $\nu_{\rm \phi}$, $\nu_{\rm r}$ and $\nu_{\rm \theta}$. The presence of explicit expressions for  $\nu_{\rm L}$ and $\nu_{\rm U}$ allows evaluation of the angular momentum $a$ of the central black hole to be made if its mass $M$ is known from alternative, independent measurements.

The underlying mechanism of LFQPOs is far from clear so there are not many models for them. What makes the situation even more intricate is that even the classification of the LFQPOs is still not complete \citep{ABC}.  The orbital frequency and the epicyclic frequencies take too high values in the vicinity of the ISCO to be associated with the LFQPOs. Several models  \citep{StellaVietriModel, Ingram_Done_model,Not_Frame_Dragging,GRO_alternative,ABC} relate the LFQPOs (the horizontal-branch QPOs in the case of Z-type neutron-star sources and the C-type QPOs in the case of black holes) to frame dragging of off-equatorial orbits and Lense-Thirring (LT) precession\footnote{We should mention for the reader, however, that this relation has been questioned in \cite{Not_Frame_Dragging}.}. The idea that LF QPOs occur due to orbital precession was proposed for the first time in the papers of \citet{StellaVietriModel} and \citet{StellaVietriMorsink} and it has later been further exploited. In the simpler models precession of single particle orbits is considered.  The more elaborate recent models proposed by \citet{PrecessingRing}, \citet{Ingram_Done_model} and \citet{Ingram_unified_model} associate the LF QPOs with the precession of thick disc which is formed by the hot inner accretion flow and which rotates as a solid body. \cite{PrecessingRing} assumed that the thick disc orbits the black hole along geodesic trajectories. In other words the precession frequency of the thick disc is given by the LT frequency of a single free particle. In the model developed by \citet{Ingram_Done_model} and \citet{Ingram_unified_model} the precession frequency of the thick disc is given by the weighted average of the LT frequencies of all test-particle circular orbits in the region that it spans.

Unlike the models for the HF QPOs those for the LF QPOs cannot be directly applied for the estimation of the mass and angular momentum of the central object. Even if we assume that LFQPOs are associated with the relativistic precession frequency in order to determine the angular momentum of the central object we need  extra information. We have to fix the radius at which a given frequency originates. Since the  LFQPOs vary in a wide interval of values it is usually accepted that they are not related to a fixed radius. What can be said for certain, however, is that according to the RP model they originate in the accretion disc, i.e. outside of the ISCO. This requirement imposes a rather loose constraint on the angular momentum of the central object. To our knowledge, it has been applied for the first time in \cite{PrecessingRing}. It turns out that for a typical BHB with $M=10M_{\odot}$ and $\nu_{\rm LF}\approx10$ Hz the lower limit for its angular momentum would be $a\gtrsim0.25$ \citep{PrecessingRing}.

In their recent paper \citet{ZdenekStuchlik} confronted several models for the HF QPOs with the requirement that they explain the extremely high value of the angular momentum of the source GRS 1915+105 that has been obtained by the continuum fitting method \citep{McClintock1}, from strong jet emission \citep{GRS_alternative} and or by the relativistic Fe-line profile fitting method \citep{GRS_Fe_line}. Only two of the models -- the $3:2$ NR model and the Keplerian $3:2$ NR model -- met this criterion. Here we confront some of the HF QPOs models considered in \citep{ZdenekStuchlik} with another criterion -- we want them to comply with a set of rather general assumptions that are met in literature concerning the nature of the central object and the origin of the LF QPOs. In the current work we focus on models which assume that $\nu_{\rm L}$ and $\nu_{\rm U}$ occur on the same orbit. The orbit is specially chosen so that the ratio of these frequencies is rational.

The aim of the presented work is not to compare the predictions of the different models for the HF QPOs for the angular momentum of the central object with values obtained by other methods (such as the Fe-line profile method or the continuum fitting method) but to search for inconsistencies between them and the constraints following form the  assumptions.

In the current paper we  confront several models for the HFQPOs enlisted below in Table~\ref{table_models} with the following assumptions:
\begin{enumerate}
\item The central object in BHBs is a Kerr black hole.
\item The (C-type) LFQPOs occur due to Lense-Thirring precession and $\nu_{\rm LF}=\nu_{\rm LT}$.
\item LFQPOs originate at (unknown) orbits outside of the ISCO -- $r_{\rm LF}\geq r_{\rm ISCO}$.
\item LFQPOs and HF QPOs do not necessarily occur on the same orbit.
\item The same mechanism is responsible for the HF QPOs of all of the observed BHBs, i.e. a single model can be applied in all cases.
\end{enumerate}
In other words, we propose a simple test for the HFQPOs models.

The suggested test requires that the mass of the BHBs is known from other, independent measurements and that two HF QPOs frequencies, $\nu_{\rm L}$ and $\nu_{\rm U}$, and at least one LFQPO -- $\nu_{\rm LF}$, have been measured for a given object. From all seven BHCs and BHBs for which HF QPOs have been observed only three are with known mass -- GRO 1655-40, XTE 1550-564 and GRS 1915+105. For all of them a pair of HF QPOs is seen. Twin QPOs occur also in the BHC  H 1743-322. There are no firm results for its mass, however. Some bounds can be found in \citep{Titarchuk_H1743_322_mass}. Their mass estimate for H 1743-322 is $13.3\pm3.2 M\odot$.

LF QPOs have been observed in all of the objects listed in this paragraph. Their parameters are listed in Table \ref{objects}.

The models are briefly described in the next section. 
They are put on test in Section \ref{Test}. As a byproduct we also propose a rough estimate of the LMXB H 1743-322. The results for the constraints on the mass of this object and the other three BHBs coming from the $3:2$ NR model are presented in Section \ref{Mass}.

\section{Models of the HF QPOs}\label{Models}
In the current study we focus mainly on two large groups of models for the HF QPOs -- the relativistic precession models and the nonlinear resonance models. The prescription for the formulas of the lower and the upper frequency of the HF QPOs given by the tidal disruption model (TD)  coincides with the one of the $2:1$ Keplerian resonance model so conclusions for these models based on the currently proposed test are the same. The warped disk model is also very popular so we will also include it in our considerations.
\begin{table*}
\centering
\begin{minipage}{140mm}
 \caption{LMXB containing a black hole for which twin HF QPOs have been observed.}
\small
\begin{tabular}{ |c|c|c|c|c|c|c|}
\hline
source          & $M/M\odot$       & $a$                    & $\nu_{LF}$, [Hz]  & $\nu_L$, [Hz]  &$\nu_U$, [Hz]& Ref.\\
\hline
GRS 1915+105    & $10.1 \pm 0.6$   & $<0.98$                & 10                          & $113 \pm 5$    &$168 \pm 3$  & 1, 2, 3, 4, 5, 6 \\
\hline
XTE 1550-564    &  $9.1 \pm 0.6$   & $0.29$\, -- \,$0.52$   & 10 (18)                         & $184 \pm 5$    &$276 \pm 3$  &
            7, 8, 9 ,10\\
\hline
GRO 1655-40     & $6.30 \pm 0.27$  & $0.65$\, -- \,$0.75$   & 28                          & $300 \pm 5$    &$450 \pm 3$  & 11, 12\\
                &  $5.31 \pm 0.07$ & $0.287$\, -- \,$0.293$ &                             &                &     &13, 14\\
\hline
H 1743-322      &  -- &    $-0.1$ -- $+0.5$                             &  22                  &   $166 \pm 5$  &  $242 \pm 3$    &15, 16, 17, 18, 19\\
\hline
     \end{tabular}
    \label{objects}
\end{minipage}
\begin{minipage}{140mm} 
References in Table \ref{Models}: 1 \citet{McClintock1}, 2 \citet{McClintock_mass_GRS}, 3 \citet{GRS_alternative}, 4 \citet{GRS_LF_10}, 5 \citet{BHBs_McClintock}, 6 \citet{Remillard_Xray_BHBs};
7 \citet{McClintock3}, 8 \citet{XTE_LF_10}, 9 \citet{XTE_LF_18}, 10 \citet{XTE_HF};
11 \citet{McClintock2}, 12 \citet{Remillard_GRO};
13 \citet{Strohmayer}, 14 \citet{GRO_alternative};
15 \citet{McClintock_H1743_322_angular}, 16 \citet{Belloni_H1743_322_LFQPOs}, 17 \citet{McClintock_H1743_322_LFQPOs}, 18 \citet{H_1743_322_HFs}, 19 \citet{Zhang_H1743_322_QPOs_last}.
\end{minipage}\\
\end{table*}

\subsubsection*{Relativistic precession (RP) models }
The frequency of the LF QPOs, $\nu_{\rm LF}$, has been related to the LT frequency $\nu_{\rm LT}$ for the first time in the relativistic precession (RP) model proposed by \citet{StellaVietriModel,StellaVietriMorsink,StellaVietriKerr}. Their model associates the upper frequency $\nu_{\rm U}$ directly with the orbital frequency $\nu_{\rm \phi}$, the lower frequency $\nu_{\rm L}$ with the periastron precession frequency $\nu_{\rm per}=\nu_{\rm \phi}-\nu_{\rm r}$ and the frequency of the LF QPOs $\nu_{\rm LF}$ -- with the nodal precession frequency which is actually the Lense-Thirring frequency $\nu_{\rm LT}=|\nu_{\rm \phi}-\nu_{\rm \theta}|$. According to the RP model, all these three QPOs are excited on the same orbit. The RP model has been applied successfully for the explanation of the $\nu_{\rm LF}-\nu_{\rm L}$ correlation that has been observed for a large number of neutron-star and black-hole sources \citep{Klis_REVIEW}. As it was mentioned in the Introduction,  \citet{Not_Frame_Dragging} found that Lense-Thirring precession could not explain the observed LF QPOs in the neutron star source Terzan 5.

\subsubsection*{Nonlinear resonance models (NRM)}
An idea that initially occurred in the papers of \citet{AlievGaltsov1, AlievGaltsov2} and later reappeared and further elaborated in the works of \citet{AbramowiczInterpreting,AbramowiczSpinEstimate, AbramowiczTheory}, relates the HF QPOs to resonances between the fundamental frequencies of motion, epicyclic and orbital, of test particles orbiting around the central object along circular geodesics in the equatorial plane\footnote{We refer the reader to \citep{AlievKerr} and \citep{Bambi2012} for applications of the NRM to Kerr and non-Kerr black holes, respectively. Nonlinear resonances occurring in the field of braneworld Kerr black holes and Kerr superspinars (Kerr naked singularities) have been studied in \citep{Braneworld} and \citep{superspinars}.}. They proposed two large groups of models -- parametric resonance models and forced resonance models.

Though very attractive due to their simplicity NRM have some significant deficiencies\footnote{Actually it seems that none of the models proposed so far can give a complete picture and explain all the observed phenomena
related to the high-frequency QPOs in black hole systems \citep{ZdenekStuchlik}. } \citep{RebuscoDifficulties}. They  do not propose an excitation mechanism for the QPOs. The origin of the coupling between the different characteristic frequencies is not clear. We should mention for the reader, however, that excitation mechanisms (more than one) have been proposed in \citep{excitation_mechanism}.

A major difficulty of the NRM is the dissonance \citep{SpinProblem, RebuscoDifficulties} between their predictions for the angular momenta of the observed BHBs and the measurements based on other methods such as: spectral continuum fitting, jet emission analysis and analysis of the profile of the $\rm K\alpha$ iron line. None of the different versions of the NRM can explain the observed angular momenta of all of the three BHCs: GRO 1655-40, XTE 1550-564 and GRS 1915+105.

\subsubsection*{Keplerian nonlinear resonance models}
Keplerian nonlinear resonance models are the first cousins of the epicyclic NRM \citep{AbramowiczInterpreting}. In the latter models the resonance occurs between the radial epicyclic frequency and the vertical one. In the former case the vertical epicyclic frequency is replaced by the orbital frequency.

\subsubsection*{Modified relativistic precession models  -- RP1 and RP2}
These versions of the precession model, RP1 and RP2\footnote{We follow the notation proposed in  \citep{ZdenekStuchlik}. } ,
are far less common than the original version proposed by \citet{StellaVietriModel} and \citet{StellaVietriMorsink}.
In both of them QPOs occur due to resonances of non-axisymmetric disc-oscillation modes with the periastron precession frequency.
In the RP1  model it is assumed that radiation is modulated by the vertical oscillations of a slightly eccentric fluid slender torus formed close to the ISCO \citep{RP1}. The RP2 model has been studied in several papers by \citet{ZdenekStuchlik,Stuchlik_mass_angular_momentum}. The model is related to Kato's work on warped-disk oscillations discussed below.

\subsubsection*{Warped-disk oscillations model (WD)}
As we have mentioned one of the drawbacks of the NRM is that it does not propose an excitation mechanism for the oscillations in the orbital motion  of the particles. An excitation mechanism is given, for example, by the warped disc model \citep{Kato2004a,Kato2004b} according to which these oscillations are resonantly excited by specific disc deformations -- warps. The WD models was applied for the estimation of the mass and the angular momentum of GRS 1915+105 in \citep{Kato2004c}. The model was generalized to include precession of the warped disk in \citep{Kato2005a},  spin-induced perturbations were included in \citep{Kato2005b}. For further development of the WD model we refer the reader to \citep{Kato2007,Kato2008}.

\subsubsection*{Tidal disruption model (TD)}
The TD model considers clumps of matter orbiting the central black hole that are deformed by tidal forces \citep{TD1,TD2,TD3}. The clumps or blobs follow perturbed circular orbits and radiate. With this model the authors were able to produce realistic light curves and even fit the high-frequency part of the power density spectrum of the LMXB XTE 1550-564 including the twin HF QPOs.

\section{The test}\label{Test}

Each of the models for the HF QPOs that are presented above gives a prescription how the upper $\nu_{\rm U}$ and the lower $\nu_{\rm L}$ frequency can be related to the dynamical frequencies --  the orbital, the two epicyclic and the precession frequencies. Their explicit form is given in Table \ref{table_models}. Explicit formulas for the fundamental frequencies -- $\nu_{\rm \phi}$, $\nu_{\rm r}$ and $\nu_{\rm \theta}$ -- can be found in Appendix \ref{appendix}. The upper and lower frequencies depend explicitly on the free parameters of the metric (for the Kerr space-time the free parameters are the angular momentum $a$ and the mass $M$ of the central object) and on the $r$ and $\theta$ coordinates of the point particle. In the case of circular orbits in the equatorial plane $r$ varies in the interval $[r_{\rm ISCO},\infty)$,  while $\theta$ is fixed, $\theta=\pi/2$.

\begin{figure*}
\centering
\begin{minipage}[c]{0.5\linewidth}
\centering \includegraphics[width=0.8\textwidth]{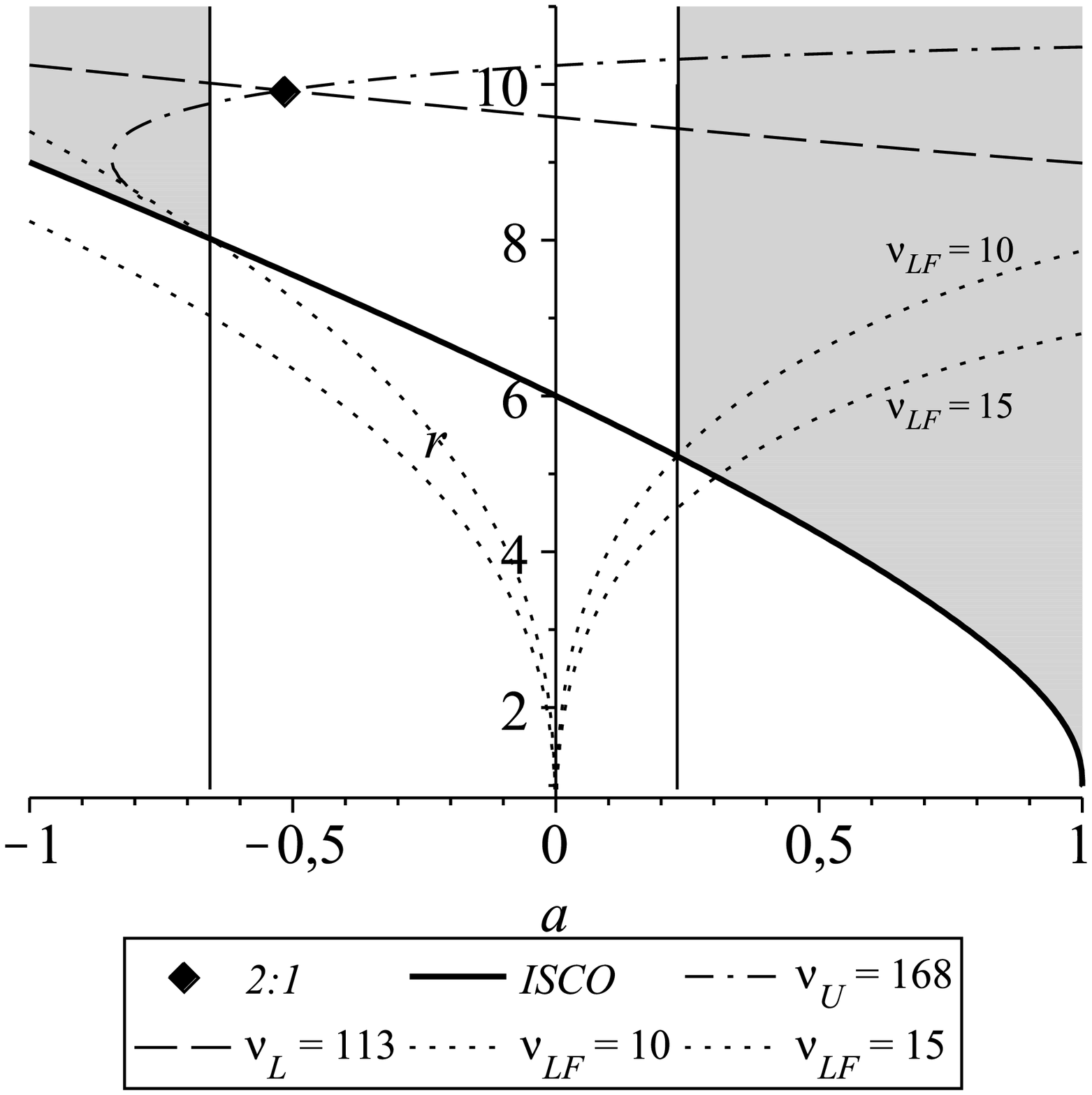}
\end{minipage}%
\begin{minipage}[c]{0.5\linewidth}
\centering \includegraphics[width=0.8\textwidth]{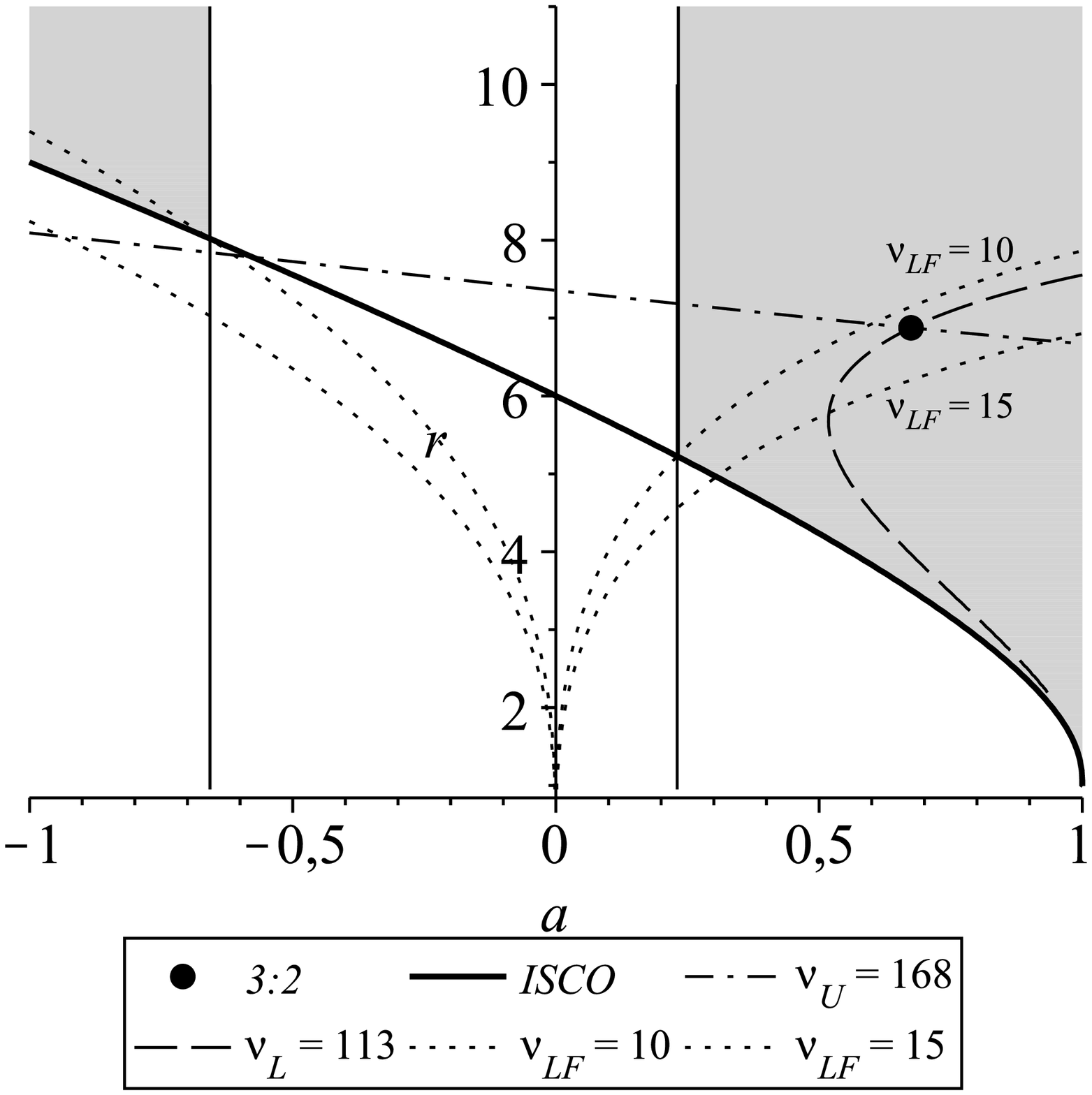}
\end{minipage}
\caption{The $r-a$ diagram representing the $2:1$ and the $3:2$ NRM. The mass and the QPO frequencies of  GRS 1915+105 have been used as an example here.}
\label{a_r}
\end{figure*}

The values of the angular momentum of the black hole and the radius of the orbit on which the HF QPOs are born can be obtained from the solution of a system of two algebraic equations $\nu_{\rm U}(r,a,M)=\nu_{\rm U}^{\rm obs}$ and $\nu_{\rm L}(r,a,M)=\nu_{\rm L}^{\rm obs}$ when the mass $M$ of object is known. 
The values of the observed frequencies $\nu_{\rm L}^{\rm obs}$ and $\nu_{\rm U}^{\rm obs}$ can be found in Table \ref{objects}. The aim of the current work is neither to give the estimates for the angular momenta of the studied objects coming from the cited models nor to compare these numbers with others obtained through different methods so no numerical values will be given explicitly here. Our aim is rather to search for possible inconsistencies between the predictions for $a$ and $r$ coming from the models for the HF QPOs and the constraints coming from the assumptions enlisted in the Introduction. Each model will just be represented by a point designated by special symbol on the angular momentum--radius diagram. The coordinates of this point come form the solutions of the above algebraic system.
\begin{table}
\caption{Models for the HF QPOs.}
\begin{center}
\begin{tabular}{ |c|c|c|}
\hline
Model &$\nu_{\rm L}$& $\nu_{\rm U}$ \\
\hline
~~3:2 & $\nu_{\rm r}$ & $\nu_{\rm \theta}$ \\
~~3:1  & $\nu_{\rm \theta}-\nu_{\rm r}$ & $\nu_{\rm \theta}$ \\
~~2:1   & $\nu_{\rm \theta}$ & $\nu_{\rm \theta}+\nu_{\rm r}$ \\
~~3:2 K  & $\nu_{\rm r}$ & $\nu_{\rm \phi}$  \\
~~3:1 K, RP  & $\nu_{\rm \phi}-\nu_{\rm r}$ & $\nu_{\rm \phi}$\\
~~2:1 K, TD  & $\nu_{\rm \phi}$ & $\nu_{\rm \phi}+\nu_{\rm r}$ \\
~~RP1   & $\nu_{\rm \phi}-\nu_{\rm r}$ & $\nu_{\rm \theta}$ \\
~~RP2   & $\nu_{\rm \phi}-\nu_{\rm r}$ & $2\nu_{\rm \phi}-\nu_{\rm \theta}$  \\
~~WD    & $2(\nu_{\rm \phi}-\nu_{\rm r})$ & $2\nu_{\rm \phi}-\nu_{\rm r}$ \\
\hline
     \end{tabular}
     \end{center}
\label{table_models}
\end{table}
Are all values of the radius and the angular momentum acceptable according to the suggested test? In other words, what positions are the special symbols that represent the different models ``allowed'' to take on the $a-r$ diagram? For each source an interval of values of the LFs has been observed. If LF is to be associated with the Lense-Thirring precession frequency then all orbits on which $\nu_{\rm LT}$ takes the values that have been observed for a given source must be outside of the ISCO. This requirement allows us to exclude some values of $a$. The tightest constraint is obtained when the maximum observed value of LF is taken. As an example, the results for GRS 1915+105 coming from two of the models, the $2:1$ and the $3:2$ NRM, are shown on Fig. \ref{a_r}. The ISCO orbit is presented by a thick solid line. To the right (left) of the thin vertical line for $a>0$ ($a<0$) the orbits on which the maximum LFQPO frequency that has been observed in GRS 1915+105, $\nu_{\rm LF}=10$ Hz, occurs is outside of the ISCO. The dashed and the dash-dotted line are the contour plots of the radii of the orbits on which $\nu_{\rm L}$ and $\nu_{\rm U}$ take the values that have been observed for GRS 1915+105. They cross at the orbit on which the corresponding resonance occurs. The allowed positions of the crossing point are in the shaded regions. Points in these regions have radii higher than ISCO and such angular momenta that the $\nu_{\rm LF}=10$ Hz orbit is outside of ISCO. On Fig. \ref{a_r} we have also added a dotted curve for $\nu_{\rm LF}=15$ Hz in order to demonstrate that higher values of LF QPOs impose tighter constraint on the angular momentum -- the curve corresponding to the higher $\nu_{\rm LF}$ crosses the ISCO line at a higher absolute value of $a$.

As it can be seen on Fig. \ref{a_r} the symbol that represents the $2:1$ model is not in the shaded region so that model defies the constraint while the $3:2$ model complies with it.
Bellow we put all of the models in Table \ref{table_models} on the test using the data for the three LMXBs: GRS 1915+105, XTE 1550-564 and GRO 1655-40.
\subsection*{GRS 1915+105}
\begin{figure*}
\centering
\begin{minipage}[c]{0.5\linewidth}
\centering \includegraphics[width=0.8\textwidth]{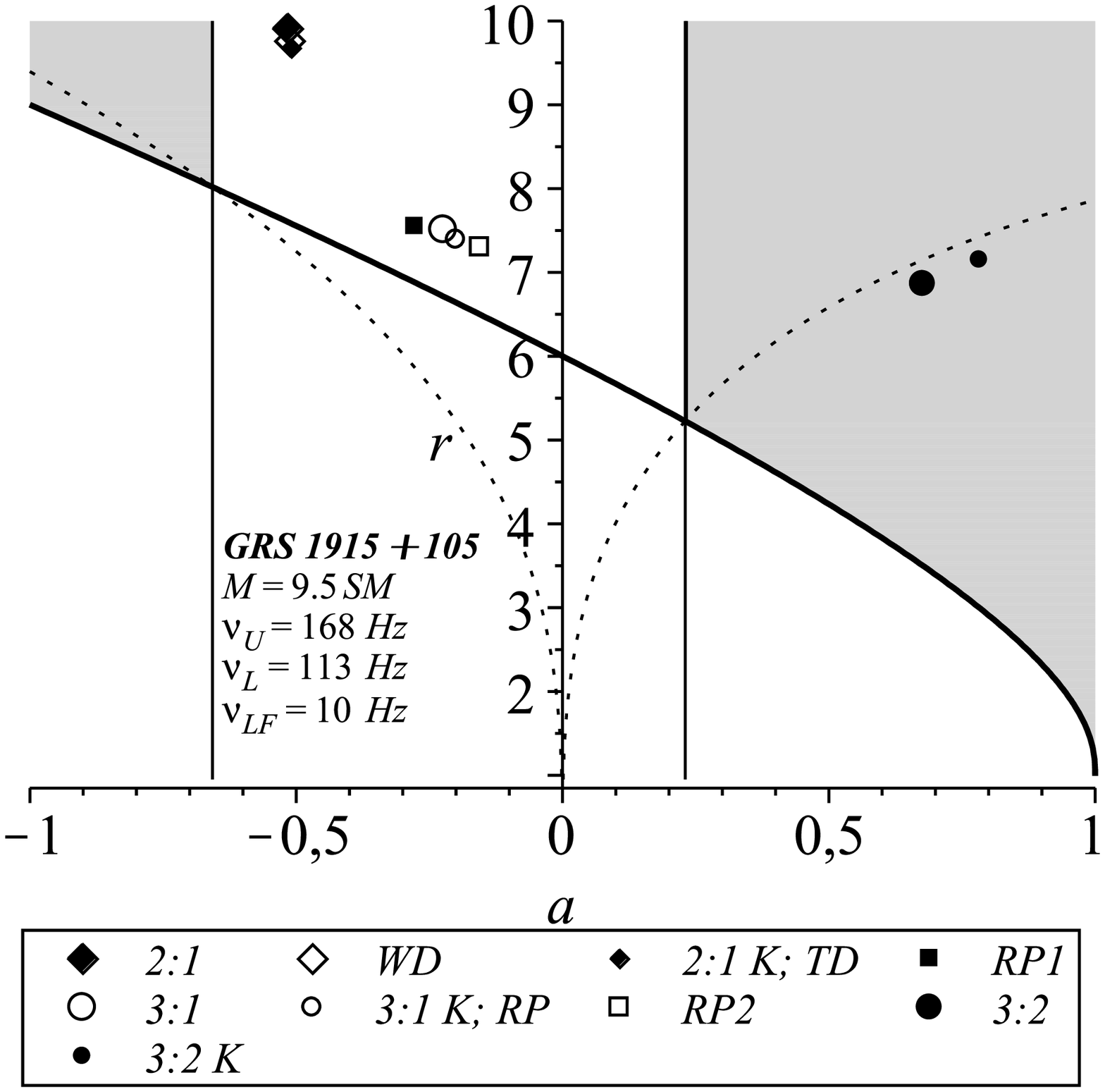}
\end{minipage}%
\begin{minipage}[c]{0.5\linewidth}
\centering \includegraphics[width=0.8\textwidth]{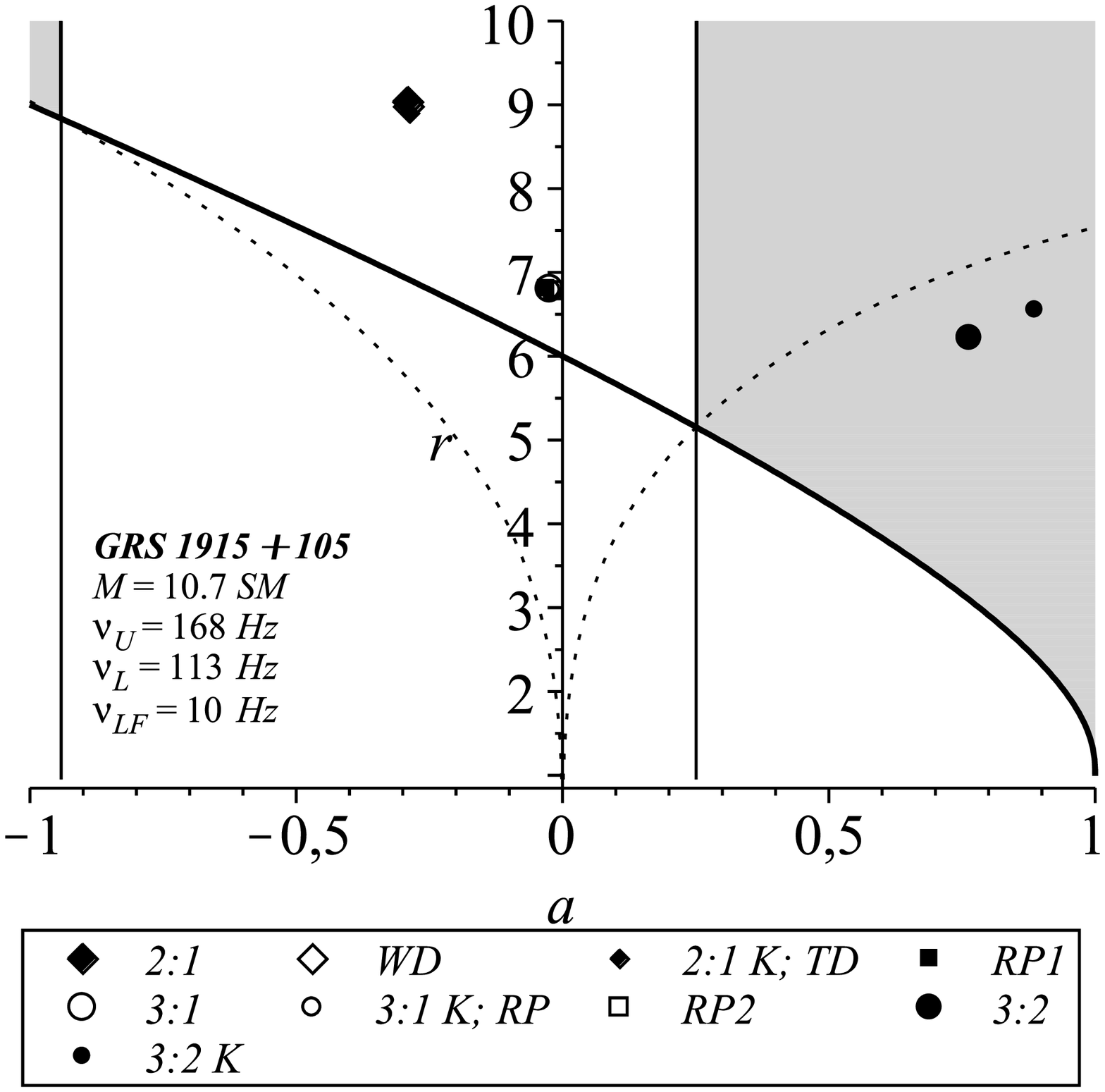}
\end{minipage}
\caption{The $r-a$ diagram presenting the results for the BHB GRS 1915+105. On the left panel the lower estimate of the mass of the black hole has been used while in the right panel -- the higher.}\label{GRS_points}
\end{figure*}

\begin{figure*}
\center
\includegraphics[width=0.44\textwidth]{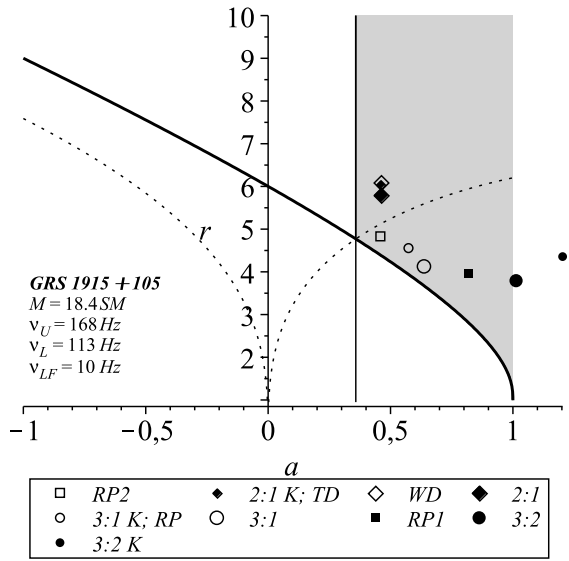}
\caption{The $r-a$ diagram for the BHB GRS 1915+105 with an alternative value for the mass -- $M/M\odot=18.4$.}
\label{GRS_points_OLD}
\end{figure*}

The results for GRS 1915+105 are shown on Fig. \ref{GRS_points}. Each model is presented by a different symbol. On the left panel the lower mass estimate has been used while on the right panel -- the upper. In both cases only two of the models comply with our constraint -- the $3:2$ NRM and the Keplerian $3:2$ NRM. The outcome of the test depends strongly on the mass of the black hole. Previously , different estimates for the mass of GRS 1915+105 were cited in literature. Its mass was estimated to be $14.0\pm4.0M\odot$ \citep{GRS_old_mass}. With this value of the mass the result from the test changes dramatically. The results are shown on Fig. \ref{GRS_points_OLD}. All of the models but the $3:2$ NRM and the Keplerian $3:2$ NRM pass the test.
Let us  check whether the $3:2$ NRM and the Keplerian $3:2$ NRM can explain the observed HF QPOs for the two other BHBs whose masses are known -- XTE 1550-564 and GRO 1655-40.
\subsection*{XTE 1550-564}
\begin{figure*}
\centering
\begin{minipage}[c]{0.5\linewidth}
\centering \includegraphics[width=0.8\textwidth]{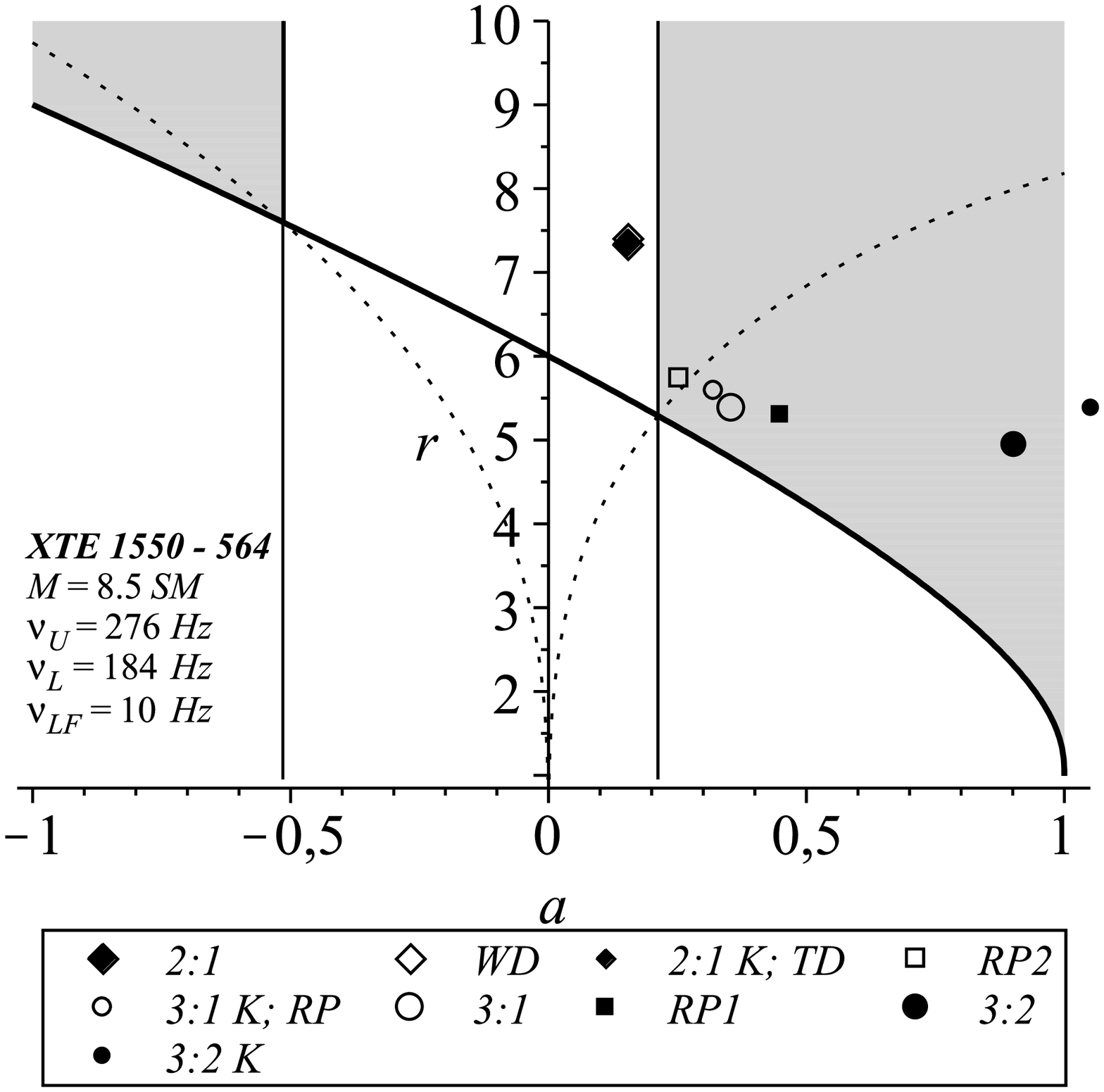}
\end{minipage}%
\begin{minipage}[c]{0.5\linewidth}
\centering \includegraphics[width=0.8\textwidth]{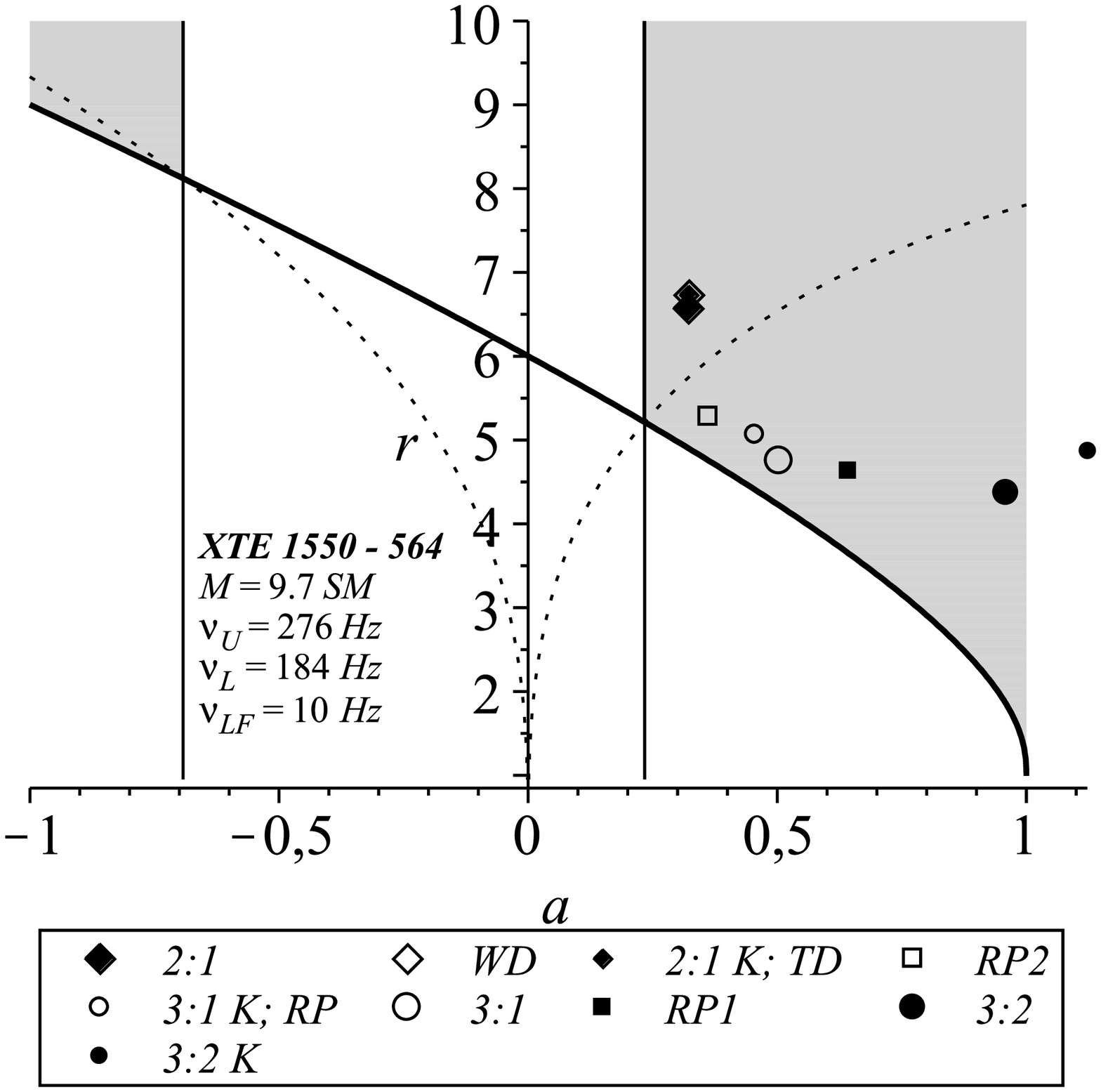}
\end{minipage}
\caption{The $r-a$ diagram for the BHB XTE 1550-564. On the left panel the lower estimate of the mass of the black hole has been used while in the right panel -- the higher. }
\label{XTE_points}
\end{figure*}

\begin{figure*}
\centering
\begin{minipage}[c]{0.5\linewidth}
\centering \includegraphics[width=0.8\textwidth]{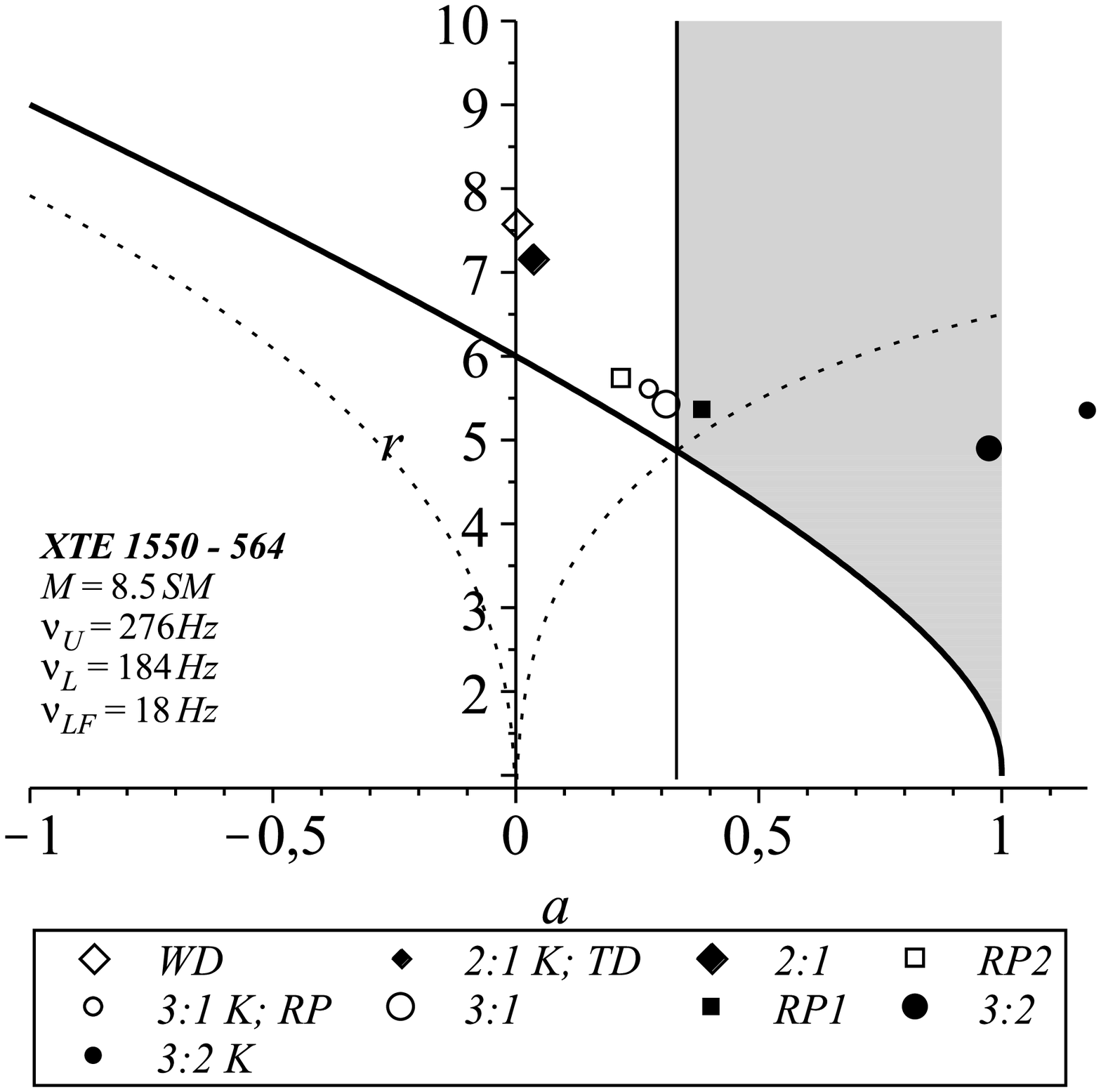}
\end{minipage}%
\begin{minipage}[c]{0.5\linewidth}
\centering \includegraphics[width=0.8\textwidth]{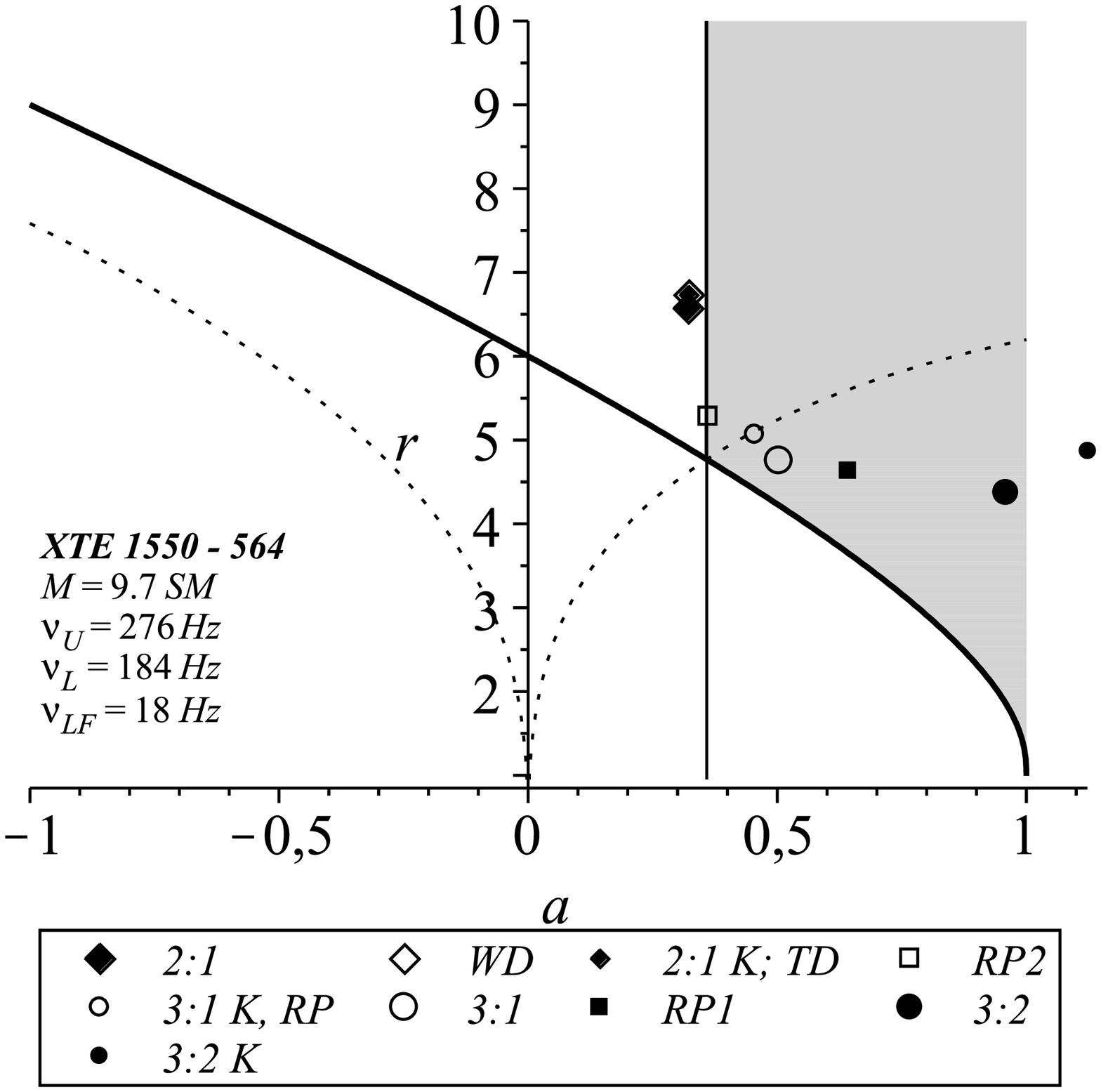}
\end{minipage}
\caption{The $r-a$ diagram for the BHB XTE 1550-564. On the left panel the lower estimate of the mass of the black hole has been used while in the right panel -- the higher. On this figure the value $\nu_{\rm LF}=18$ has been used. }
\label{XTE_points_alternative}
\end{figure*}

 The $r-a$ diagram for the BHB XTE 1550-564 is presented on Fig. \ref{XTE_points}. With both masses the $3:2$ NRM is in the shaded region. The  Keplerian $3:2$ NRM is in conflict with the proposed constraints since it predicts angular momenta higher that $a=1$. Data for the LF QPOs of XTE 1550-564 can be found also in  \citep{XTE_LF_18}. Their type (A, B or C), however, is not determined in this paper. The LF QPOs of this source have been further discussed in \citep{XTE_LF_18_futher}. If we speculate that the highest value for $\nu_{\rm LF}$ cited in Table~1 of \citep{XTE_LF_18} -- $18$ Hz -- can be related with the Lense-Thirring precession frequency we can obtain an even tighter constraint for the models. The results are given in Fig. \ref{XTE_points_alternative}. When the lower (higher) mass is used only two (six) of the 11 models that we have considered pass the test.

\subsection*{GRO 1655-40}
\begin{figure*}
\centering
\begin{minipage}[c]{0.5\linewidth}
\centering \includegraphics[width=0.8\textwidth]{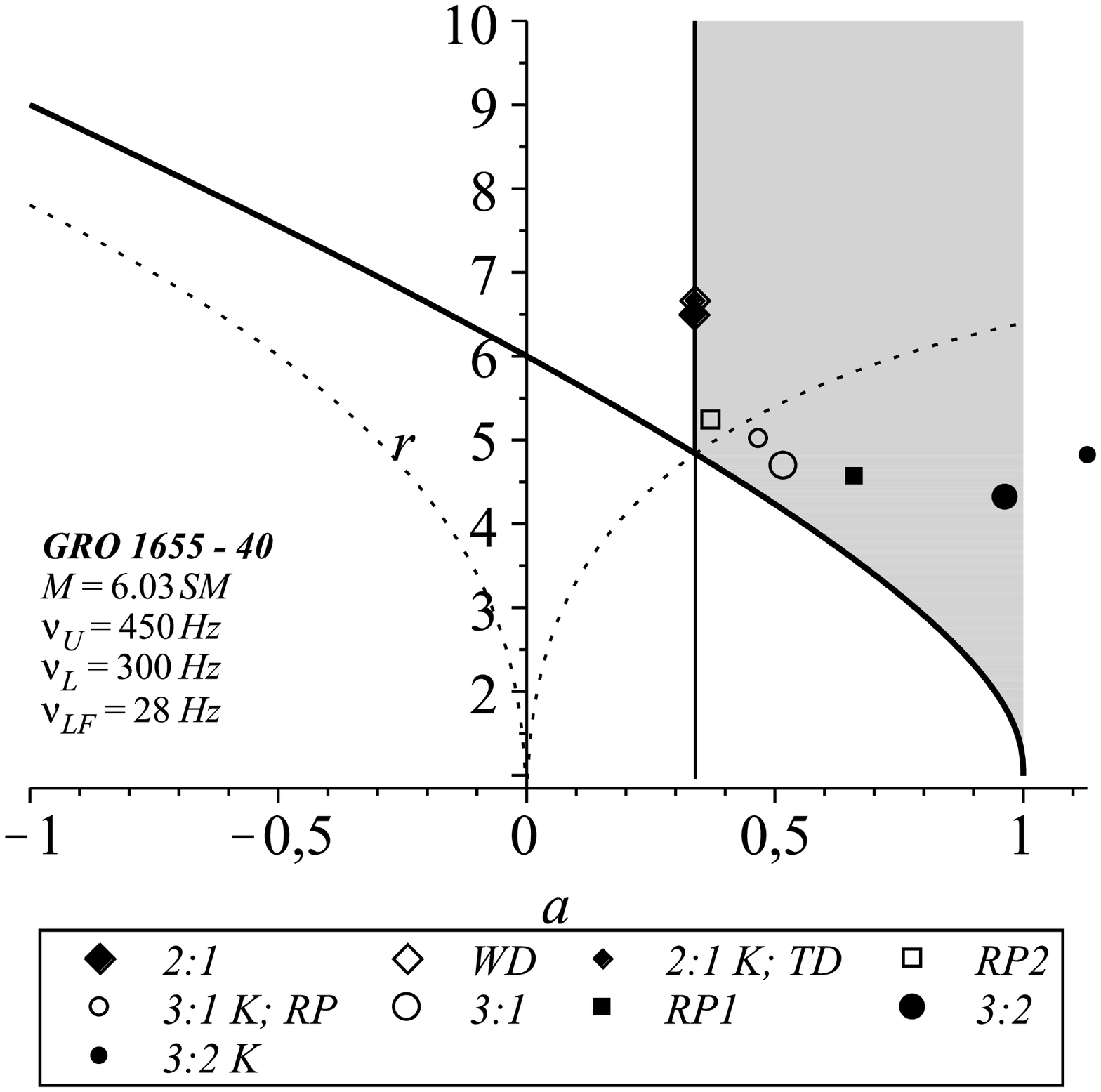}
\end{minipage}%
\begin{minipage}[c]{0.5\linewidth}
\centering \includegraphics[width=0.8\textwidth]{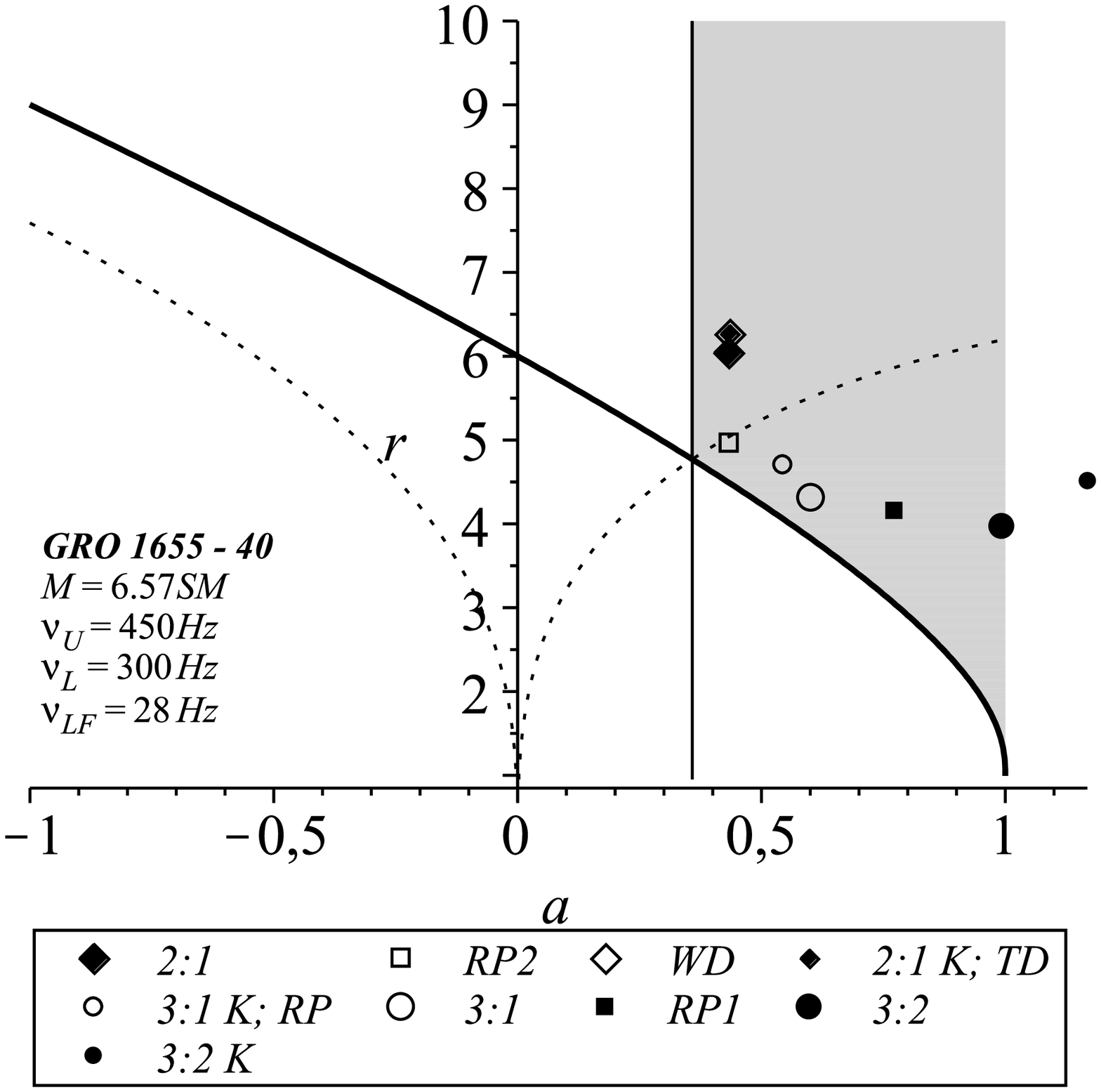}
\end{minipage}
\caption{The $r-a$ diagram for the BHB GRO 1655-40. On the left panel the lower estimate of the mass of the black hole has been used while in the right panel -- the higher. }
\label{GRO_points}
\end{figure*}

The results from the test for that object are presented on Fig. \ref{GRO_points}. As in the previous cases, on the left panel the lower mass estimate has been used while on the right panel -- the upper. What can be immediately seen is that the data excludes the possibility of counterrotating accretion disc (there is no shaded region for negative values of $a$). With both values of the mass the test excludes only the  Keplerian $3:2$ NRM  since it predicts angular momenta higher that $a=1$ which is in conflict with our first assumption (The assumptions are enlisted in the Introduction.).

\begin{figure*}
\centering
\begin{minipage}[c]{0.5\linewidth}
\centering \includegraphics[width=0.8\textwidth]{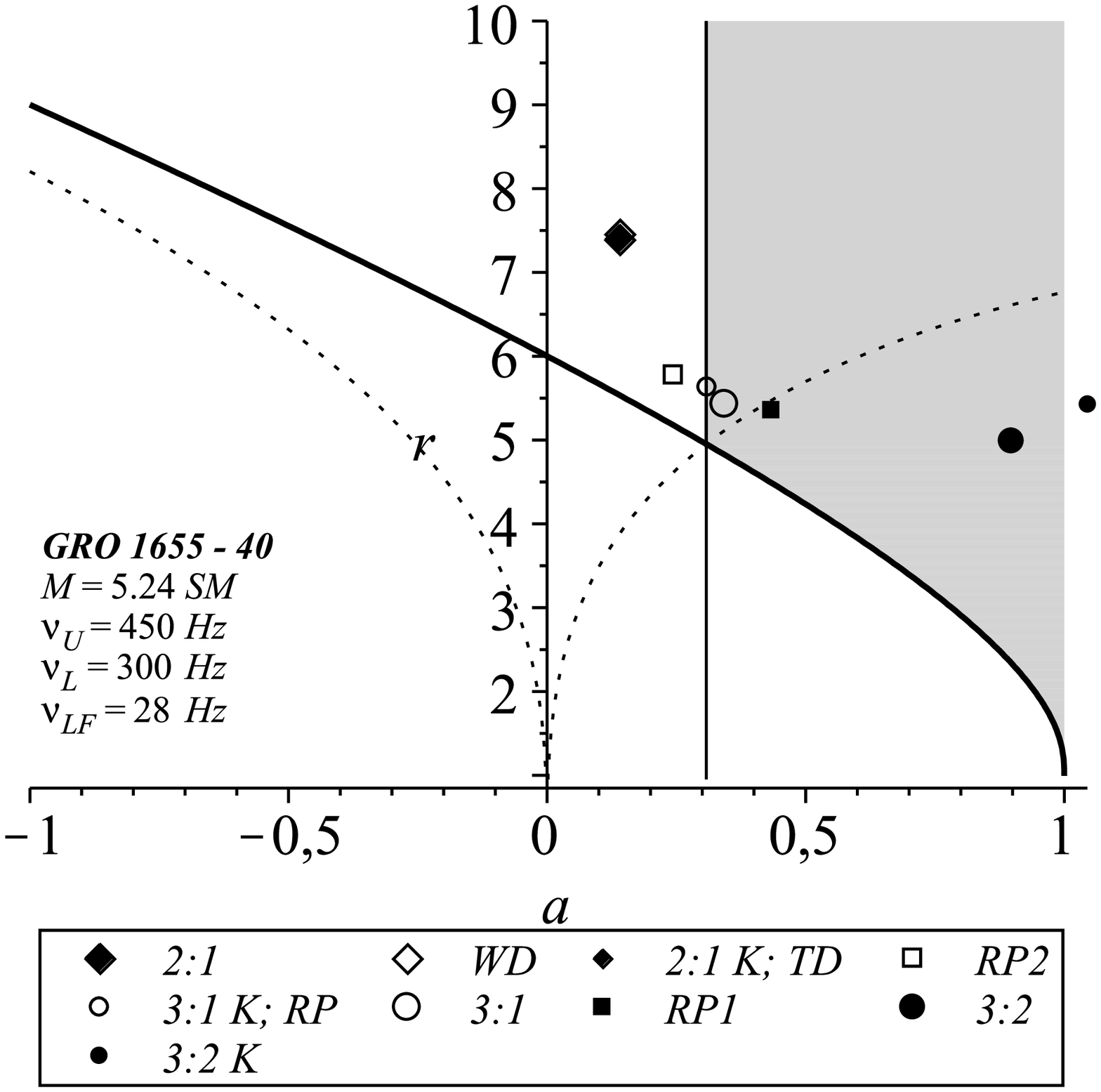}
\end{minipage}%
\begin{minipage}[c]{0.5\linewidth}
\centering \includegraphics[width=0.8\textwidth]{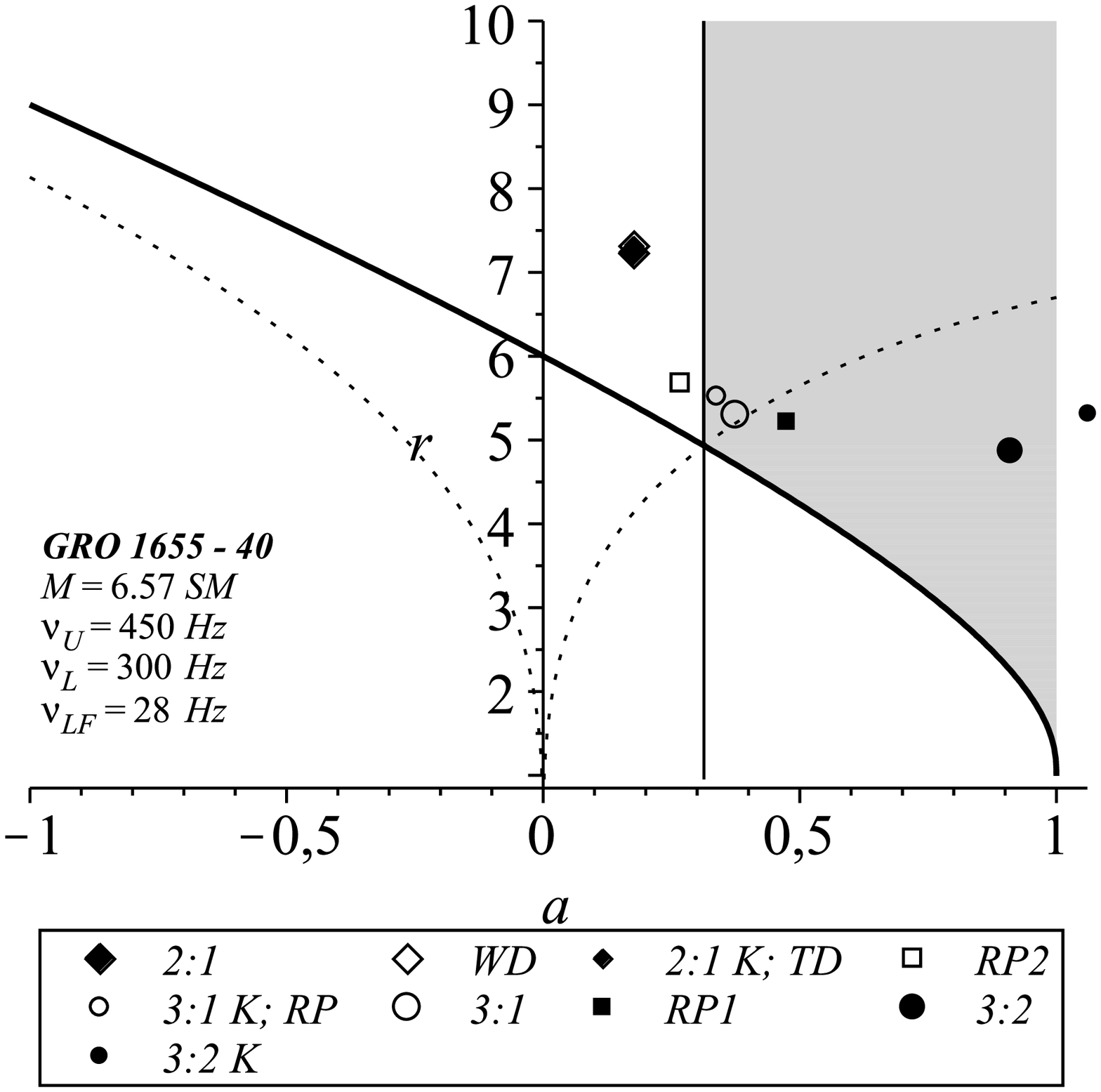}
\end{minipage}
\caption{The $r-a$ diagram for the BHB GRO 1655-40. On the left panel the lower estimate of the mass of the black hole has been used while in the right panel -- the higher. On this figure the alternative estimate for the mass of GRO 1655-40 given in \citep{GRO_alternative} has been used. }
\label{GRO_points_alternative}
\end{figure*}

For the results presented on Fig. \ref{GRO_points} we have used mass estimate which does not come from QPOs data. For completeness we have repeated the test using the alternative mass estimate for the same object given in \citet{GRO_alternative}. The latter mass estimate comes from QPOs data.
It appears that the lower masses used in Fig. \ref{GRO_points_alternative} are more restrictive and fewer models pass the test.

\section{Mass bounds}\label{Mass}
It turns out that the proposed test, though rather loose, allows us to single out a model for the HF QPOs -- namely the $3:2$ NRM.
From all the models that we have studied only this model can explain the HF QPOs of all three BHCs, which is in agreement with assumption number five,  and is not in conflict with the constraints following from the other assumptions. Once the model  which complies with all other assumptions is found it can be applied to constrain the mass of a BHC. In the previous secton the value of the mass $M$ was taken from other measurements and then one of the models for the HF QPOs was used for the evaluation of the angular momentum $a$. In order to find the constraints of the mass that a given model for the HF QPOs applies we follow an inverse procedure -- for given $a$ the mass $M$ is evaluated.

 Let us describe the procedure for the case of prograde orbits ($a>0$) first. If $M$ is increased the position of the symbol which represents the $3:2$ NRM on the $a-r$ diagram is shifted towards higher angular momenta\footnote{The same is true for all of the models considered in this paper since the frequencies are inversely proportional to the mass of the black hole.}. Hence, taking into account that for the Kerr black hole cannot be higher than $a=1$ one can put an upper bound on the mass $M$. What about a lower bound? Inversely, as $M$ is decreased the crossing point is shifted to the left . Its abscissa, however, is constraint from below by the vertical line, which gives us the required lower bound on the mass $M$. It must be taken into account, however, that the position of the vertical line also varies with $M$. Similar procedure is followed in the case of retrograde orbits ($a<0$).

The mass bounds obtained with this procedure can be compared to the reference mass bounds -- those given in Table \ref{objects}. The results are presented in Table \ref{table2}. In the first two columns of the table there are two lines for each object. The upper line contains the values for retrograde orbits, while the lower -- for prograde. In the third column the referential values of the masses are given for convenience. The values which are in agreement with the referential ones are in bold. As it can be seen, the mass constraints coming from retrograde orbits are systematically lower than the literature values. Agreement is found for positive angular momenta in all three of the cases.

Now, as a by product, we can apply the same recipe to constrain the mass of the low-mass X-ray binary H 1743-322. The mass of this BHC has not been measured yet but a pair of HF QPOs with stable $3:2$ ratio of the frequencies are observed in its power spectrum. The results for the constraints on its mass are also given in Table \ref{table2}. The mass interval obtained here partially overlaps with the one given in \citep{Titarchuk_H1743_322_mass}. Our predictions for the angular momentum for this object are also in agreement with the constraints given in \citep{McClintock_H1743_322_angular}, $a = 0.2\pm0.3$.
\begin{table}
\centering
\caption{Bounds on the angular momentum $a$ and the mass $M$ of the four microquasars based on the 3:2 NRM.  }
\begin{tabular}{ |c|c|c|c|}
\hline
source                        &  a                                    & $M/M\odot$ & ref $M/M\odot$ \\
\hline
GRS 1915+105                  &$-1.0$\, -- \,$-0.20$                  & $3.1$ -- $4.6$ & $9.5$ -- $10.7$\\
                              &$\textbf{0.16}$\, -- \,$\textbf{1.0}$  & $\textbf{5.8}$ -- $\textbf{17.6}$ &  \\
\hline
XTE 1550-564                  &$-1$\, -- \,$-0.12$                    &$1.9$ -- $3.0$ & $8.5$ -- $9.7$\\
                              &$\textbf{0.10}$\, -- \,$\textbf{1.0}$  &$\textbf{3.5}$ --  $\textbf{11.0}$ &\\
\hline
GRO 1655-40                   &$-1$\, -- \,$-0.22$                    &$1.2$ -- $1.7$ &$6.03$ -- $6.57$\\
                              &$\textbf{0.17}$\, -- \,$\textbf{1.0}$  &$\textbf{2.2}$ -- $\textbf{6.8}$ & $5.24$ -- $5.38$\\
\hline
H 1743-322                    &$-1$\, -- \,$-0.31$                    & $2.1$ -- $2.9$   & $10.1$ -- $16.5$ \\
                              &$\textbf{0.22}$\, -- \,$\textbf{1.0}$                    & $4.0$ -- $ 11.4$ &\\
\hline
     \end{tabular}
     \label{table2}
\end{table}

\section{Conclusion}
In the current paper we propose a simple test of the models for the HF QPOs. The proposed test is based on five rather general assumptions
concerning the nature of the central object in BHBs and the mechanism for the generation of the LF QPOs observed in the PDS of such objects. The test was applied to the eleven models presented in Table \ref{table_models}. The tightest constraints come from the BHB GRS 1915+105 when its new mass bounds \citep{McClintock_mass_GRS} are used. This object discards all but two of the studied models -- the $3:2$ NRM and the Keplerian the $3:2$ NRM. The constraints coming form the other two BHBs that were considered -- XTE 1550-564 and GRO 1655-40 -- are not so tight. However both of these objects discard the Keplerian the $3:2$ NRM since it requires $a>1$ -- a value which is in conflict with the assumption that the central object is a Kerr black hole. Finally,
since one of our basic assumptions, number 5, is that one and the same model can explain the  HF QPOs of all of the observed BHBs then only one of the models that we studied -- the $3:2$ NRM, is not discarded. Once we could single out the $3:2$ NRM, as a byproduct, we applied it to put loose constraints on the LMXB H 1743-322. For counterrotating accretion disc our mass bounds are $2.1\leq M/M\odot \leq 2.9$, for corotating -- $4.0\leq M/M\odot \leq 11.4$.

\section*{Acknowledgments}
This work was partially supported by the Bulgarian National Science Fund under Grant No DMU 03/6. The author would like to thank prof. Stoytcho Yazadjiev for reading the manuscript and Dr. Sava Donkov for the encouragement.

\appendix

\section{Fundamental frequencies}\label{appendix}
The explicit form of the orbital frequency $\nu_{\rm \phi}$ and the two epicyclic frequencies -- the radial $\nu_r$ and the vertical $\nu_{\theta}$ -- for the Kerr black hole can be found, for example, in \citep{AlievKerr}
\begin{equation}
\nu_{\rm \phi} =\left({1\over 2\pi}\right)\frac{ M^{1/2}}{ r^{3/2} \pm a M^{1/2}}\,\,,
\label{orbf}
\end{equation}
\begin{equation}
\nu_{r}^2 = \nu_{\rm \phi}^2\, \left( 1-\frac{6 M}{r} -\frac{3
a^2}{r^2} \pm \, 8 a {M^{1/2}\over r^{3/2}}
\right),
\label{kerrradf}
\end{equation}
\begin{equation}
\nu_{\theta}^2= \nu_{\rm \phi}^2\, \left(1
+\frac{3 a^2}{r^2} \mp \, 4 a {M^{1/2}\over r^{3/2}} \right).
\label{kerraxif}
\end{equation}

\end{document}